\documentclass[12pt]{article}

\usepackage[margin=1in]{geometry}  
\usepackage{graphicx} 
\usepackage{dcolumn}
\usepackage{hyperref}
\usepackage{cite}
\usepackage{url}

 \usepackage{lscape}

\title{A Decade of News Forum Interactions: Threaded Conversations, Signed Votes, and Topical Tags}

\author{
Emma Fraxanet$^{1,2}$\thanks{Corresponding author: emma.fraxanet@bsc.es} \\
Vicenç Gómez$^{2}$ \\
Andreas Kaltenbrunner$^{2,3}$ \\
Max Pellert$^{1}$\thanks{Corresponding author: max.pellert@bsc.es}
}

\date{}

\begin{document}
\maketitle

\begin{center}
{\small
$^{1}$Barcelona Supercomputing Center (BSC-CNS), Barcelona, Spain \\
$^{2}$Department of Engineering, Universitat Pompeu Fabra, Barcelona, Spain \\
$^{3}$Ethical Technologies and Connectivity for Humanity Research Center, Universitat Oberta de Catalunya, Barcelona, Spain}
\end{center}

\begin{abstract}
We present a large-scale, longitudinal dataset capturing user activity on the online platform of DerStandard, a major Austrian newspaper. The dataset spans ten years (2013-2022) and includes over 75 million user comments, more than 400 million votes, and detailed metadata on articles and user interactions. It provides structured conversation threads, explicit up- and downvotes of user comments and editorial topic labels, enabling rich analyses of online discourse while preserving user privacy.
To ensure this privacy, all persistent identifiers are anonymized using salted hash functions, and the raw comment texts are not publicly shared. Instead, we release pre-computed vector representations derived from a state-of-the-art embedding model.
The dataset supports research on discussion dynamics, network structures, and semantic analyses in the mid-resourced language German, offering a reusable resource across computational social science and related fields.
\end{abstract}

\section*{Background and Summary}

Social media have become a defining aspect of the first quarter of the $21^{st}$ century. For researchers, social media platforms are attractive data sources, as they offer access to large-scale, naturally occurring, and continuous streams of data on human communication and interaction. However, we increasingly find that straightforward access to such data is restricted, with many commercial platforms limiting or shutting down public APIs~\cite{murtfeldt2024rip}. Moreover, mainstream social platforms known for their discussions oriented at specific topics, such as political events or science, can quickly undergo rapid user migration or shifts in topical focus, as has been witnessed with Twitter/X~\cite{bisbee2025vibes}. In this context, online discussion forums attached to news media sites represent a valuable and relatively stable alternative. In prior work, news-based comment sections have been shown to influence readers' perceptions of article quality~\cite{naab2020comments, prochazka2018effects}, to shape opinions~\cite{ziegele2014creates}, and to serve as counterpublic spaces that challenge dominant narratives~\cite{toepfl2015public}. These platforms typically combine structured comment threads, editorial context, and high-quality moderated textual content, making them particularly well suited for longitudinal and community-level analyses~\cite{bacaksizlar2023group, ha2025dynamics}. 

We present a large-scale longitudinal dataset that provides user discussions on the online platform of DerStandard, an Austrian newspaper. Covering the full decade from 2013 to 2022, the dataset includes over 75 million user timestamped comments, more than 400 million up- and downvotes on these comments, and detailed metadata on the news articles, including editorial topic tags. A detailed overview of the file structure and metadata fields is provided in Figure~\ref{fig:data_records}.

DerStandard is an Austrian print newspaper that was founded in 1988. As early as 1995, it went online, marking the first appearance of a major German-language newspaper on the web, according to its own claim. Its early adoption of digital community features, starting with chat rooms and evolving into discussion spaces for registered users below the news articles, has led to a highly active user forum. Today, DerStandard remains one of the most widely visited online news platforms in Austria, reaching several million unique users per month according to audience measurements by the Austrian Web Analysis (ÖWA). Prior research has also identified DerStandard as one of the most trusted traditional news sources among young Austrian audiences and as a central platform for online news consumption~\cite{russmann2020news}.

\begin{figure}
    \centering
    \includegraphics[width=0.95\linewidth]{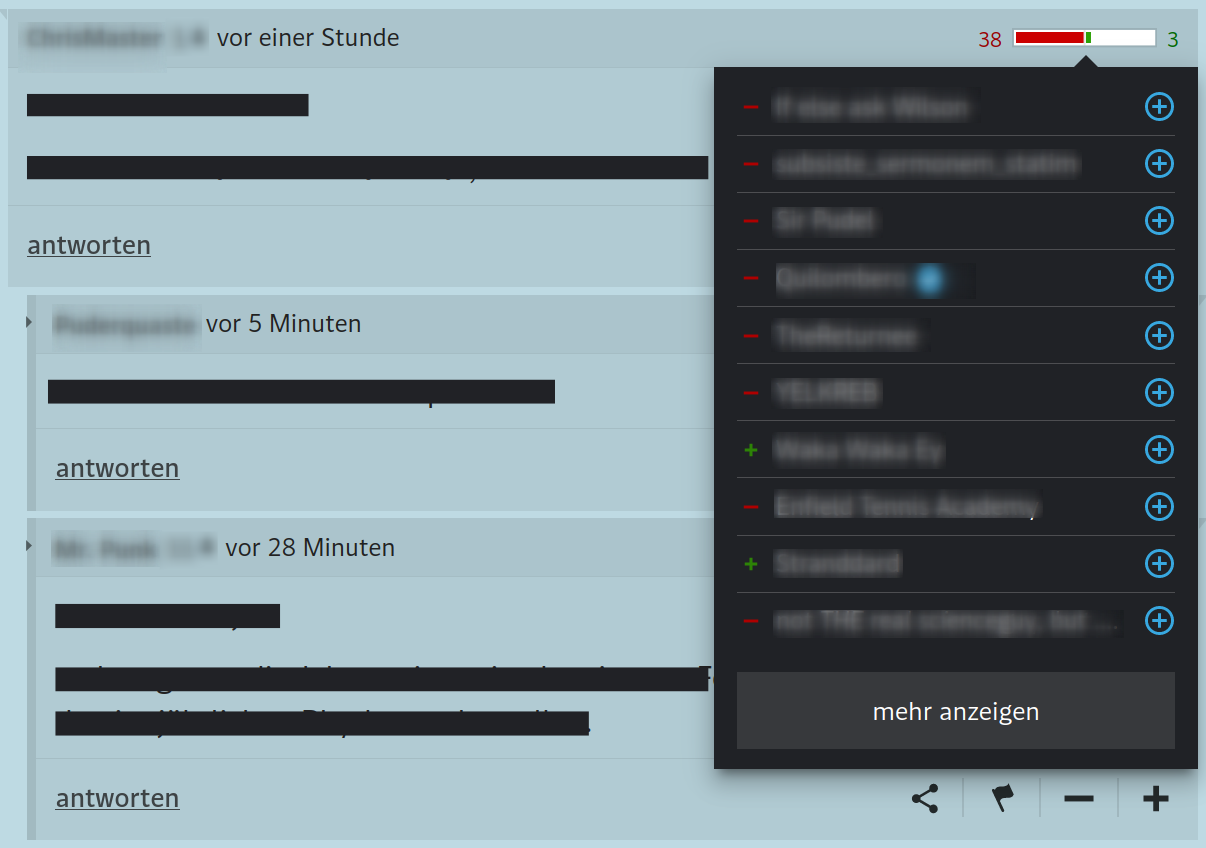}
    \caption{\textbf{Example of the DerStandard comment platform interface.} The screenshot shows a user comment with its associated voting summary in the upper right corner, indicating the number of upvotes (green) and downvotes (red). By hovering over the bar, users can see a breakdown of who voted and how (up- or downvote). Comments are displayed in threaded format with timestamps and reply links. We blurred user names and blacked out comment bodies and titles in this example. Users can also share and flag a comment as inappropriate (by clicking on the share and flag icons, respectively).}
    \label{fig:example}
\end{figure}

Users can post comments under articles, vote on others' comments, and engage in structured discussion threads (see Figure~\ref{fig:example} for an example of the interface). In contrast to comment sections on social media platforms, which are often shaped by commercial incentives, comment sections on news websites are frequently conceived as spaces intended to foster public discussion and the exchange of opinions among readers~\cite{russmann2025management}. Studies of Austrian news forum community management highlight that DerStandard maintains one of the largest moderation teams among Austrian news outlets and actively invests in fostering structured and civil discussion in its comment sections~\cite{russmann2025management}.  Moderation is performed using semi-automated systems that have been developed in collaboration with external research institutions such as the Austrian Research Institute for Artificial Intelligence (OFAI) \cite{schabus-skowron-2018-academic}, resulting in the minimization of low-quality text content on the platform. 

Survey evidence from three representative Austrian waves in 2020 (Waves 4–6; total n = 3002)~\cite{niederkrotenthaler2022mental} reported by Pellert et al.~\cite[p. 2–6 in SI]{pellertValidatingDailySocial2022} indicates that DerStandard users are, on average, more male, younger, and more highly educated than the Austrian population; however, this evidence reflects a limited time window and does not provide a longitudinal demographic characterization of the full ten-year dataset.

To protect user privacy, we anonymize all persistent identifiers such as user and comment IDs using a salted cryptographic hash (see Section~\ref{sec:methods}), and the raw text of comments is not publicly distributed by us. To respect these constraints while still enabling research that requires the use of text, we pre-computed vector representations for each of the more than 75 million texts in the dataset that we publicly share. We use a state-of-the-art, specialised open embedding model ``KaLM-embedding-multilingual-mini-v1''(\url{https://huggingface.co/HIT-TMG/KaLM-embedding-multilingual-mini-v1}) from the Hugging Face Model Hub.

With this dataset, we extend beyond the predominantly (U.S.) English-centric focus of many large-scale online datasets by providing a rich data source centered on German, a mid-resourced language spoken by around 100 million people worldwide, in the specific context of Austria.
The period covered by our data includes numerous contentious political events, both in Austria and the world in general. Despite being a small European country, Austria is often considered a model for broader sociopolitical trends in Germany and other Western European countries. Its location in Central Europe and historical role as a crossing point between Eastern, Western, and Southern Europe give it shared characteristics with many of its neighboring countries.

The data uniquely combines several features. First, it has a strong longitudinal scope, covering a full decade in the first quarter of the 21st century. Second, the platform's active voting mechanism on comments provides explicit information on the \textit{sign} of user interactions (up- or downvotes). Third, we retrieved detailed metadata on the comment as well as the article level from the platform, including editorial topic tags that organize articles into a three-level topical hierarchy. These features make the dataset particularly well-suited for advancing research on online human behavior, especially through the analysis of network structures~\cite{candellone2025negative}, context-dependent interactions~\cite{barbera2015tweeting}, thread conversations~\cite{aragon2017thread}, and temporal dynamics~\cite{kaltenbrunner2007description}. Particularly, explicit agreement and disagreement interactions between users are usually not available to researchers, and proxy measures, such as Wikipedia edits \cite{maniuBuildingSignedNetwork2011} or Reddit replies~\cite{pougue2021debagreement}, have often been used instead to infer positivity or negativity of interactions. The dataset also includes metadata on deleted users, enabling studies of user churn and moderation effects. Importantly, this dataset provides unique opportunities to study large-scale online interactions and can be readily extended to include future years by public webscraping. 

Because comments and votes are linked to their parent articles, the topical structure allows to subset discussions by policy area, political actor, or specific events, and to track the salience and dynamics of topics over time. This makes the dataset suitable for integration with external longitudinal datasets commonly used in political science and public opinion research. For example, the temporal structure and fine-grained topics related to specific parties can be aligned with policy agenda datasets such as the Comparative Agendas Project~\cite{baumgartner2019comparative}, parliamentary corpora such as ParlaMint~\cite{parlamint}, or party positioning data from the Chapel Hill Expert Survey~\cite{rovny2025ches} to examine how political agendas are reflected in online discussions. Similarly, the dataset can be combined with repeated cross-sectional surveys measuring public attitudes, such as the European Social Survey (\url{https://www.europeansocialsurvey.org/}), European Values Study (\url{https://europeanvaluesstudy.eu/}), or Eurobarometer (\url{https://europa.eu/eurobarometer/screen/home}), allowing researchers to relate shifts in public opinion to patterns of engagement and polarization in online discussions. In particular, the availability of explicit agreement and disagreement signals through votes enables the study of interactional forms of polarization, which can provide empirical indicators of affective polarization between groups of users. Considering this dataset covers events such as the coronavirus pandemic, issue-specific panel datasets such as the Austrian Corona Panel Project~\cite{P5YJ0O_2020} can help compare online discussion dynamics and changes in public attitudes during major societal events.

Data from DerStandard has previously been included in publicly released annotated datasets. While these datasets are very valuable because of their manual annotations and released text content, they do not offer the possibility to study the structural properties of user interactions or discussion threads. The One Million Post Corpus~\cite{schabusOneMillionPosts2017} features over one million user comments, with a subset manually annotated for content moderation tasks. GERMS-AT\cite{krenn2024germs} contains comments annotated for varying degrees of sexism and misogyny, developed for the GermEval2024 shared task on sexism classification at the Conference on Natural Language Processing (KONVENS) (See \href{https://ofai.github.io/GermEval2024-GerMS/}{https://ofai.github.io/GermEval2024-GerMS/}). The availability of these datasets highlights the platform's ongoing relevance for research. The other publicly released data sets complement our work by providing annotated subsets that support supervised learning and evaluation tasks. We describe in detail the relation between the datasets, including overlap and integration challenges, in Section~\ref{sec:tech_val}.

The data shared in this article has already been used in three separate studies. The first work provided near real-time monitoring of the affective expressions of the entire DerStandard community during COVID-19~\cite{pellertDashboardSentimentAustrian2020a}. Second, it has been shown that the results of sentiment analysis on users' comments on DerStandard closely track the dynamics of explicitly expressed mood by users in a traditional survey that was run on the platform in the same time period~\cite{pellertValidatingDailySocial2022}. Third, information on vote interactions was used in order to extract signed networks of users and identify the main divisions on the platform. This analysis allowed to pinpoint issues that reinforce societal fault lines and contribute to polarization, as well as topics, such as COVID-19, that spark online conflict without strictly following these dividing lines~\cite{10.1093/pnasnexus/pgae276}. In Section~\ref{sec:tech_val}, we include the mapping of user IDs to the two factions identified in that study, enabling researchers to analyze behavior and interactions through the lens of long-term ideological alignment.

We provide technical validations for several aspects of our dataset. We confirm internal consistency of the provided data files and quantify potentially problematic cases such as deleted users or mismatched timestamps. In addition, we validate the embeddings that we share by comparing semantic similarity scores across reply structures and topic categories. We find that the embeddings encode both conversational structure and thematic coherence, making them suitable for semantic analysis for many research questions without having to access the full text. 

Altogether, "A Decade of DerStandard" provides a rare combination of scale, depth, and structure with the added benefit of a non-English setting. It is a valuable resource for research domains such as computational social science, network analysis, and natural language processing research.

\section*{Methods}\label{sec:methods}

\begin{figure}
    \centering
    \includegraphics[width=0.82\linewidth]{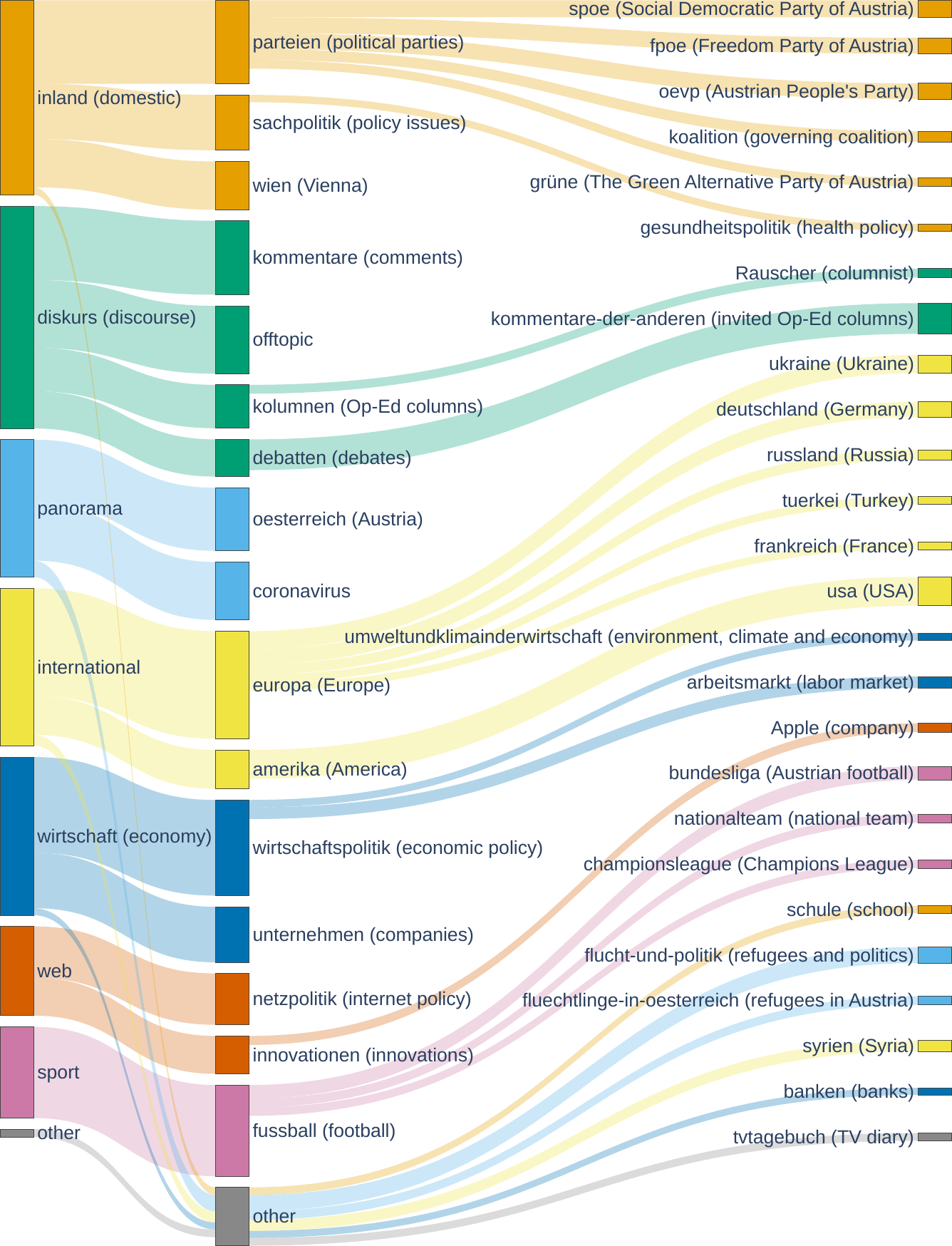}
    \caption{\textbf{Most discussed topics across DerStandard's internal tag hierarchy.} The diagram shows the internal tag hierarchy used by DerStandard editors to classify articles, highlighting the tags associated with the highest total number of user comments between 2013 and 2022. Flows between columns indicate how general categories are structured into more specific topics, proportional to the number of comments going to each subtopic. Node size indicates the volume of comments in each category. We color nodes according to the top-level category that they belong to. The "other' category represents any top- or mid-level tags that are not in the diagram due to having significantly lower comment counts than the ones displayed.}
    \label{fig:enter-label}
\end{figure}

\paragraph{\textbf{Data Acquisition.}} We retrieved all information in the dataset by sending standard HTML requests to the public platform at \url{derstandard.at}. We used a combination of scripts, building on the \textit{curl} library. We took care to balance the load on the platform with waiting times and added information to the user header that identified our requests as automated. All the information that we retrieved is publicly available to any visitor of the site, i.e., accessible without being logged into a user account on the platform. We downloaded raw HTML files and used XPath with \textit{xidel} as well as \textit{jq} for JSON processing to extract relevant objects. We also scraped article metadata. For example, we extracted the article-associated topic tags annotated by DerStandard (see Figure~\ref{fig:enter-label}). We converted all extracted data fields to a common TSV format. The scripts used to retrieve the raw HTML pages and perform the data extraction are included in the \texttt{scripts/} folder of the public dataset repository.

\paragraph{\textbf{Anonymization.}}To protect user privacy while preserving relational links within the dataset, we anonymized all original identifiers of users and comments derived from the platform with new unique values created through a salted cryptographic hashing function. We used the BLAKE2s hashing algorithm~\cite{aumasson2013blake2} with an 8-byte digest and a fixed, undisclosed salt value. Original identifiers were converted to strings, concatenated with the salt, and hashed to produce consistent but irreversible anonymized values. This approach ensures that the same anonymized, hashed ID always maps to the same original identifier across all files, allowing for linkage (e.g., from comments to users or from votes to voters) without exposing sensitive information. Article identifiers were not anonymized, so they can be used to access the corresponding articles directly on the DerStandard platform (by appending the \textit{Article\_Id} at the end of https://derstandard.at/story/\textit{Article\_Id}).

\begin{figure}[hb]
    \centering
    \includegraphics[width=0.75\linewidth]{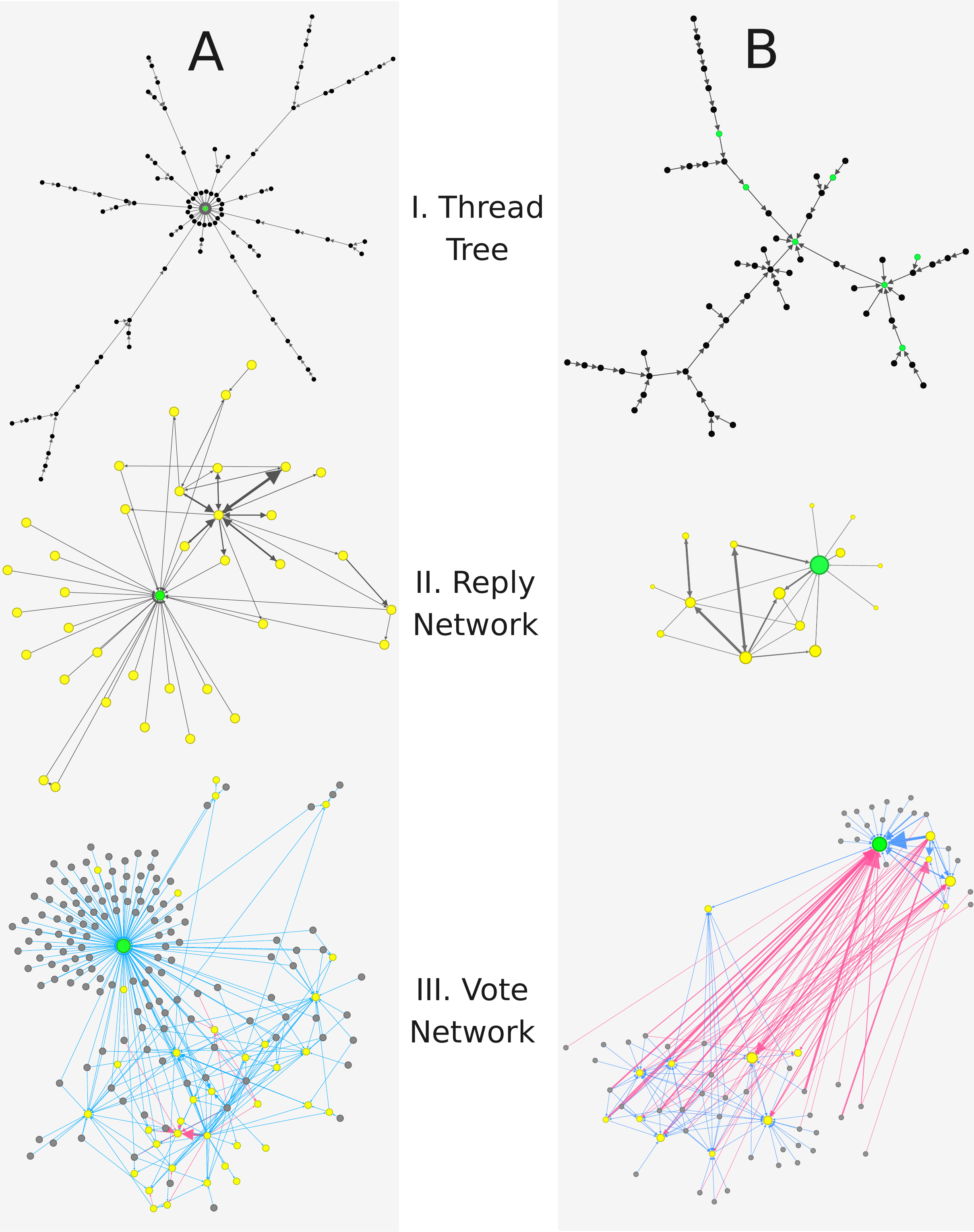}
    \caption{\textbf{Example forum conversations from three different perspectives.} We illustrate two thread discussions on the platform using three visualizations: (I) the thread structure, showing the hierarchical relation between comments (with comments by the original comment author in green); (II) the reply network among users (highlighting the original post author in green); and (III) the vote network between users, including both commenting users (in yellow) and users who only voted (in gray), as well as the original author (in green). Example A shows a larger, mostly positive discussion, while Example B features a smaller, more decentralized discussion with a polarized voting pattern.}
    \label{fig:thread_networks}
\end{figure}

\paragraph{\textbf{Preparation of Additional Metadata: Thread Information and Users Dataset.}}To support research that requires computationally intensive operations over the full dataset, we pre-compute and provide several forms of aggregated metadata, for example, discussion thread characteristics. A thread refers to a structured sequence of comments starting from a top-level post and including all its nested replies. For each comment, we identify its thread origin parent, i.e., the top-level comment in the thread to which it belongs, as well as its thread depth, defined as the number of reply steps separating it from the root comment. These fields, together with the direct parent, allow a faster reconstruction of thread structures and the analysis of conversational hierarchies. For specific research questions, this provides information that allows researchers to do an initial sampling of the data (for example, by considering threads of a certain depth only). We show two example thread discussions and their corresponding reply and vote networks in Figure~\ref{fig:thread_networks}.

We create a comprehensive users file containing one row per user who has either posted a comment or cast at least one vote. This file includes user-level aggregates such as the total number of comments written, total upvotes and downvotes received and given, and timestamps of first and last recorded activity (voting or commenting). These aggregated user profiles facilitate analyses of user behavior and engagement over time, with the opportunity of pre-selecting a subset of engaged and active users.

\paragraph{\textbf{Text Embeddings.}}We generate embeddings of each comment in our dataset to include in the public release instead of the full text to protect the privacy of their content and avoid maintaining a public copy of, in the meanwhile, possibly deleted comments. 

Recent work has shown that certain semantic attributes, recognizable entities, sensitive information, or short phrases can sometimes be inferred from embeddings~\cite{morris2023text, jha2025harnessing, huang2024transferable}, particularly when the embedding model is known. Some approaches require access to both pretrained and fine-tuned embeddings to exploit their differences~\cite{hayet2022invernet}, while others rely on generative models trained on auxiliary data to produce paraphrased versions of the input~\cite{li-etal-2023-sentence}. However, based on the current literature, no published work has demonstrated scalable, high-fidelity verbatim reconstruction of full original texts from plain fixed sentence embeddings like those released on our dataset. Furthermore, under our anonymization procedures, the embeddings cannot be used to re-identify individual users.

We provide dense vector representations with 896 dimensions for each comment using the ``KaLM-Embedding''~\cite{hu2025kalm} approach. We selected this specialised class of models for the task because they generally perform well in benchmarks such as the Massive Text Embedding Benchmark \cite{muennighoffMTEBMassiveText2023}. Specifically, we used the "KaLM-embedding-multilingual-mini-v1" (\url{https://huggingface.co/HIT-TMG/KaLM-embedding-multilingual-mini-v1}) model from the Hugging Face Model Hub (Commit ID of the exact model version used: \href{https://huggingface.co/HIT-TMG/KaLM-embedding-multilingual-mini-v1/tree/8a82a0cd2b322b91723e252486f7cce6fd8ac9d3}{8a82a0cd2b322b91723e252486f7cce6fd8ac9d3}). The manageable size of around 2 GB of the model and smaller model complexity compared to many other approaches, allowed us to create a CPU-only workflow on the high-performance computing infrastructure available to us. We split the corpus of comments into arbitrary chunks (24 chunks for each of the 120 months of our data) to create a highly parallelizable setup for model inference.

\section*{Data Records}

\begin{figure}[h]
    \centering
    \includegraphics[width=0.75\linewidth]{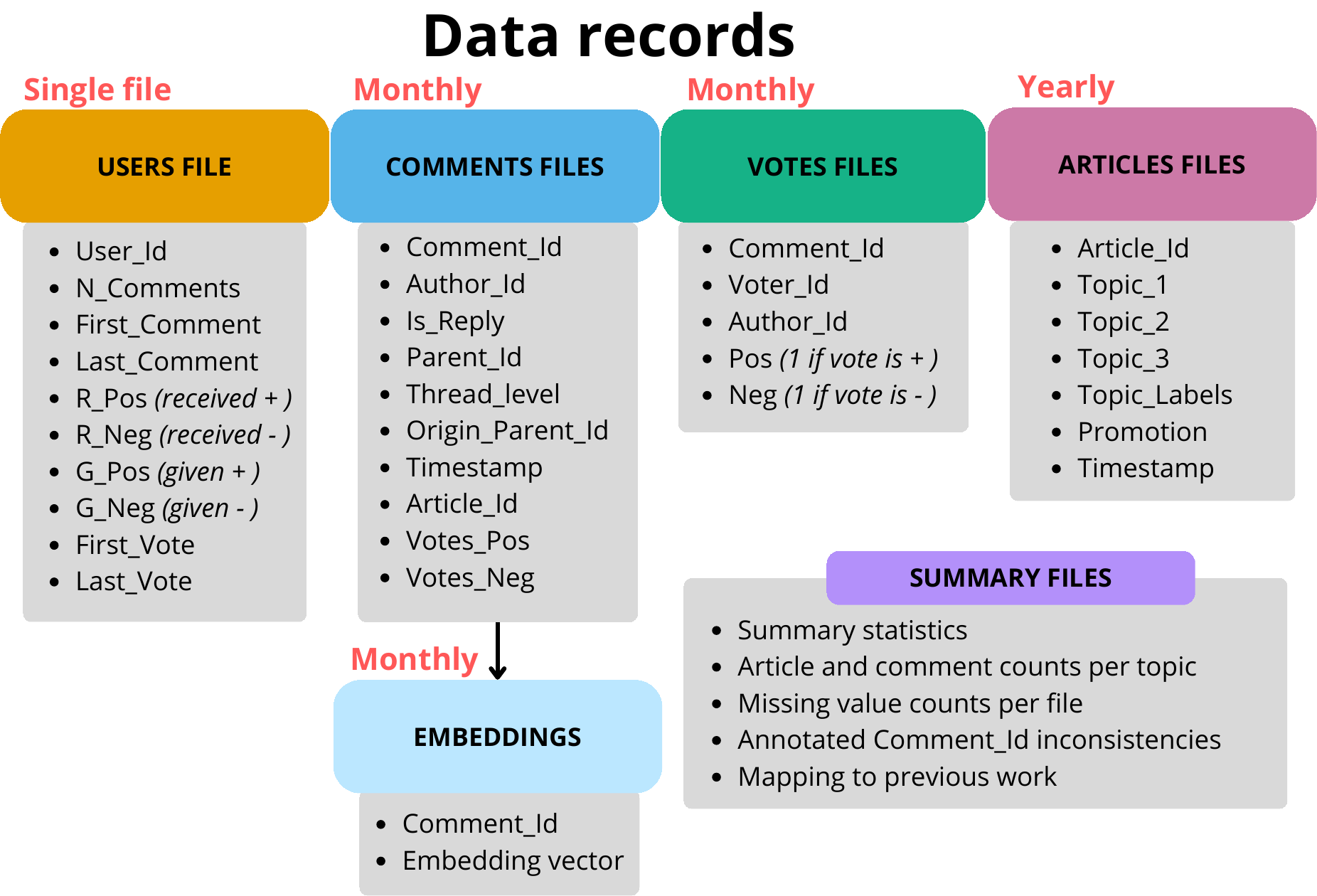}
    \caption{\textbf{Overview of the data records included in the DerStandard dataset.} The dataset is structured into several files based on record type and temporal granularity. A single \texttt{Users file} contains anonymized metadata per user, including activity statistics and voting behavior. \textit{Comments}, \textit{Votes}, and \textit{Embeddings} are provided as monthly files, capturing all user-generated content, up- and downvotes, and pre-computed text embeddings, respectively. \texttt{Articles files} are aggregated yearly and include metadata such as timestamps, and up to three editorial topic labels per article. \texttt{Summary Files} contains auxiliary information, including aggregate statistics, data quality annotations, and mappings to previous work. All variable names coincide with the column header in the files. Additional explanations of some variables are provided in parentheses.}
    \label{fig:data_records}
\end{figure}

The full dataset is shared as a Dataset inside the BSC Dataverse \textit{A Decade of DerStandard Forum Interactions}~\cite{RVQGUG_2026}, organized into separate directories by data type and temporal resolution. These include user-level information, comment-level content, voting behavior, and article metadata, as well as pre-computed text embeddings. The files are stored in compressed TSV format for accessibility and ease of use. Except for the user file, all data is provided in either monthly or yearly resolution to facilitate efficient loading and period-specific analysis. An overview of the different data components and their variables is shown in Figure~\ref{fig:data_records}. The folders in the dataset, including data components and other summary and helper files, are the following:

\begin{itemize}
    \item Articles Files (Yearly)
    \item Comments Files (Monthly)
    \item Users Files (Single file)
    \item Votes Files (Monthly)
    \item Embeddings (Monthly)
    \item Summary Files, addressing data inconsistencies, derived summary metadata and mapping to previous work annotations.
    \item Scripts, containing basic example code for accessing and processing the data (e.g. to retrieve summary files)
\end{itemize}

Relationships between records across files can be established through shared identifiers. For instance, comment files link to the user file via the \texttt{Author\_Id} and \texttt{User\_Id} fields, and each comment is associated with the corresponding news article through the \texttt{Article\_Id}. All users—whether they appear as authors of comments, voters or both—are included in the user file. In addition, every comment can be linked to its corresponding embedding vector using the \texttt{Comment\_Id}. As detailed in Section~\ref{sec:methods}, all user and comment identifications are anonymized for privacy protection. A separate file with a list of comments with small inconsistencies is also provided in the folder. Refer to Section~\ref{sec:tech_val} for a detailed explanation of these inconsistencies and the presence of empty fields (i.e., "NA" values).

In total, the dataset covers over $75$ million comments, $400$ million votes, and more than $580$ thousand articles, spanning ten years from 2013 to 2022. A brief summary of dataset size and user activity is presented in Table~\ref{table:stats}.

\begin{table}[!t]
\centering
\caption{\textbf{Summary statistics of the DerStandard dataset.} We provide key metrics on the number of articles, user comments, votes, and users between 2013 and 2022, including averages, standard deviations, and activity indicators.}
\label{table:stats}
\begin{tabular}{@{}l|l|r@{}}
                  & \textbf{Statistic}                         & \textbf{Value}        \\ \hline
\textbf{Articles} & Total                             & 586 942       \\
                  & Per year (average)                & 58 694      \\
                  & Per year (stdDev)           & 10 726      \\
                  & Number of unique Topic\_1 labels  & 25           \\
                  & Number of unique Topic\_2 labels  & 305          \\
                  & Number of unique Topic\_3 labels  & 1 722         \\ \hline
\textbf{Comments}    & Total                             & 75 644 850     \\
                  & Per month (average)               & 630 374    \\
                  & Per month (stdDev)          & 17 183       \\
                  & Number of threads                 & 21 957 527     \\
                  & Number of replies                 & ($\sim71\%$) 53 617 134     \\ \hline
\textbf{Embeddings} & Vector dimension                 & 896 \\ \hline
\textbf{Votes}    & Total                             & 412 511 165     \\
                  & Total positive                    & ($\sim77\%$) 319 154 131       \\
                  & Per month (average)               & 3 437 590       \\
                  & Per month (stdDev)          & 1 000 560      \\ \hline
\textbf{Users}    & Total                             & 247 863       \\
                  & \% That voted more than 10 times  & $\sim$52\%   \\
                  & \% That commented more than 10 times & $\sim$36\%   \\
                  & \% Active for at least 1 year     & $\sim$44\%   \\
                  & \% Active for 10 years            & $\sim$0.45\%
\end{tabular}
\end{table}

\section*{Technical Validation}\label{sec:tech_val}

\paragraph{\textbf{Quality Validations of Embeddings.}}To validate that the pre-computed text embeddings accurately capture the semantic structure of user comments, we conducted a series of validation analyses. These analyses confirm that the embeddings preserve meaningful relationships between comments at both conversational and topical levels.

\begin{figure}[b!]
    \centering
    \includegraphics[width=0.55\linewidth]{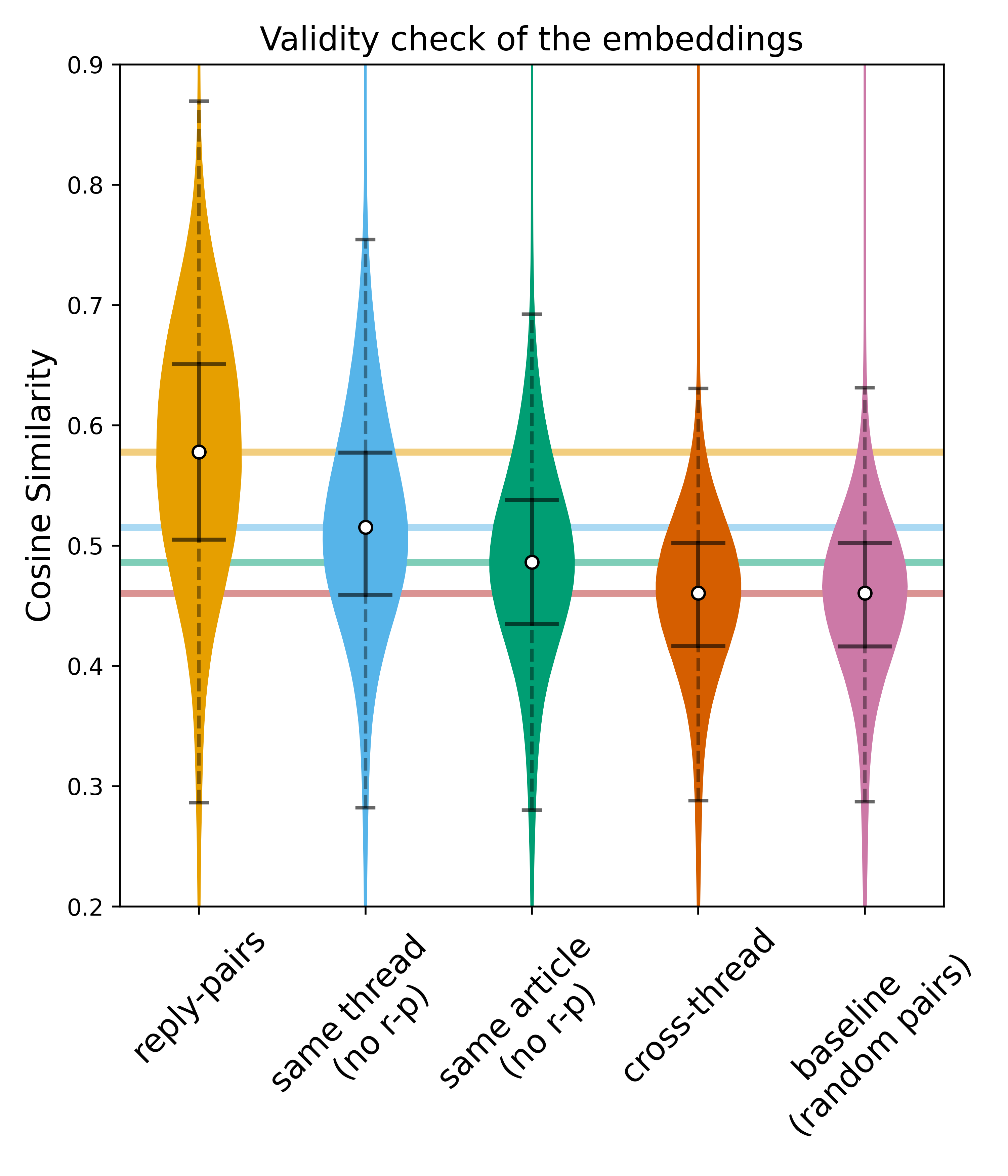}
    \caption{\textbf{Validation of comment embeddings based on discussion structure.}
Cosine similarity distributions for different types of comment pairs. 
Similarity is highest for direct reply pairs, followed by comments within the same thread, and then comments under the same article (excluding direct replies). Cross-thread and random pairs show the lowest similarity, indicating that semantic proximity in the embeddings aligns with structural proximity in discussions.}
    \label{fig:similarity_rp}
\end{figure}

We first evaluate whether the cosine similarity between comment embeddings reflects structural proximity in discussion threads (Figure~\ref{fig:similarity_rp}), decreasing from direct replies (most proximal) to pairs from different threads and articles (most distant). Despite considerable variability in similarity scores across comment pairs, the average similarities decrease consistently with structural proximity: highest for reply pairs, followed by non-reply comments within the same thread, comments under the same article, and lowest for cross-thread and random pairs, which serve as a baseline. These results indicate that the embeddings capture semantic properties of comments that are consistent with the structure of discussion threads.

\begin{figure}[b!]
    \centering
    \includegraphics[width=0.98\linewidth]{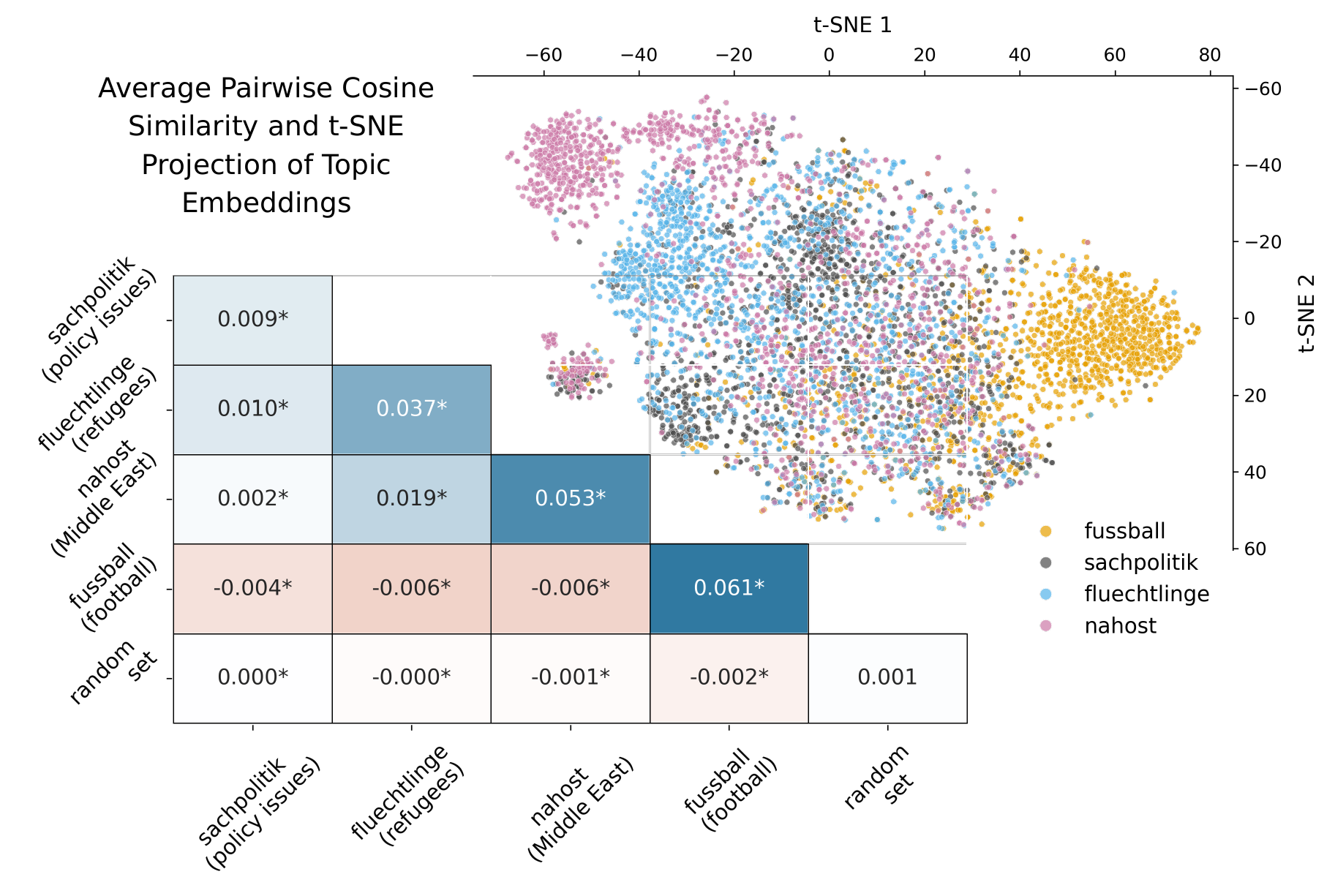}
  \caption{\textbf{Semantic coherence of topic-specific comment embeddings.} 
\textbf{Left:} Average pairwise cosine similarity among 1M pairs of comment embeddings within and across selected editorial topics, compared against a random baseline. Higher within-topic similarity (diagonal) indicates that the embeddings capture topic-specific semantic content, particularly for focused topics such as \textit{football}. Cosine similarity is computed after subtracting the mean embedding of $1.2$M random comments to remove shared semantic structure. Asterisks (*) denote significance based on a two-sample t-test against the random self-similarity distribution ($p < 0.001$). Blue (red) in the matrix boxes indicates a positive (negative) similarity.
\textbf{Right:} Two-dimensional t-SNE projection (after PCA reduction to 50 dimensions) of 5,660 comment embeddings across four selected topics. Clear clustering is observed for focused topics, and spatial proximity between topics reflects the similarity patterns observed in the left panel.}
    \label{fig:similarity_topics}
\end{figure}

We next examine whether comments associated with the same editorial topic exhibit semantic cohesion (Figure~\ref{fig:similarity_topics}). We compute cosine similarity after subtracting the mean embedding of $1.2$M random comments to remove shared semantic structure. We also compare all topic sets with a random set and use the self-similarity of the random set as a baseline distribution to compare significance (based on a two-sample t-test). Within-topic similarities (diagonal) are consistently higher than cross-topic similarities, with more focused topics such as \textit{refugees} and \textit{football} showing stronger internal coherence, while broader categories like \textit{policy issues} display slightly lower cohesion. When comparing different topics, we observe that semantically related topics (e.g., \textit{refugees} and \textit{policy issues} or \textit{refugees} and \textit{Middle East}) exhibit higher cross-topic similarity, whereas unrelated topics (e.g., \textit{football} and \textit{Middle East}) show lower similarity levels, even below the comparison to the random set.

As a qualitative illustration of our embedding validation, the t-SNE projection on the right of Figure~\ref{fig:similarity_topics} provides a view of topic clustering in the embedding space. We first reduced the 896-dimensional comment embeddings to 50 principal components using Principal Component Analysis (PCA), retaining approximately 24\% of the variance, and then applied t-SNE with a perplexity parameter of 50. The resulting visualization shows that comments associated with each topic form distinguishable clusters, with \textit{football} forming the most compact and semantically distinct cluster.

Together, these analyses illustrate that the KaLM-based embeddings capture semantic relationships that reflect both conversational and topical structures, suggesting their potential usefulness for tasks such as topic classification, semantic retrieval, and comment clustering.

\paragraph{\textbf{Mapping to Previous DerStandard Datasets.}}DerStandard has previously been used to create and share datasets supporting a variety of research goals.
The One Million Post Corpus released by OFAI~\cite{schabusOneMillionPosts2017} contains over one million user comments, with a manually annotated subset targeting categories relevant to content moderation.
By matching exact timestamps, we identified $1,415$ of these annotated comments in our dataset. Unmatched comments may fall outside our time frame or may have been removed by users or moderators, as the original annotations focused primarily on inappropriate or discriminatory content.

The GERMS-AT dataset~\cite{krenn2024germs}, developed for the GermEval 2024 shared task on sexism classification at the Conference on Natural Language Processing (KONVENS) (See \href{https://ofai.github.io/GermEval2024-GerMS/}{https://ofai.github.io/GermEval2024-GerMS/}), includes approximately $8,000$ DerStandard comments annotated for varying degrees of sexism and misogyny. Although based on the same platform, GERMS-AT does not include metadata or identifiers that would enable linkage with our dataset.

While integration with previously annotated corpora is limited, we provide a user-level mapping (\texttt{summary\_files/faction\_users}) to the two main factions identified in prior work on affective polarization using DerStandard data~\cite{10.1093/pnasnexus/pgae276}. This study applied the FAULTANA pipeline to detect disagreement fault lines on the platform by constructing a signed user-user network based on patterns of positive and negative interactions (i.e., comment votes). The method identifies an optimal partition into user factions by minimizing the number of frustrated edges—those that contradict the assumption of within-group agreement and across-group disagreement.

Using this approach, two main factions were detected, reflecting a persistent division in the user community over the 2014–2021 period. We include these faction labels as a user-level attribute for the $14,817$ users categorized in the study, with approximately 62\% belonging to the larger faction. These labels provide a structural grounding for future research on ideological alignment, polarization dynamics, and cross-cutting issue detection. Note that the mapping is static and reflects long-term alignment; it does not account for potential shifts in user position over time due to future events or evolving platform dynamics.

\paragraph{\textbf{Internal Consistency.}}Missing values occur in certain fields due to user deletions. We keep this data because it might be useful for research that studies user abandonment or migration in discussion-based platforms. On average, $8.46\%$ of comment authors are missing (\texttt{Author\_Id} = NA) in the comments files, with a standard deviation of $3.78\%$ across months. Similarly, $8.33$($\pm3.66$)\% of author fields are missing in the votes files, and $7.27$($\pm3.35$)\% of voter identifiers are absent. 

To ensure internal coherence of the dataset, we conducted a series of validation steps across all files and identifiers. We first verified that all \texttt{Comment\_Id} values appearing in the votes files are also present in the comments files, ensuring referential integrity between user votes and the associated content. 

We highlight a few anomalies and provide an explanation on how these issues have been dealt with in the published dataset: We keep the data and provide a separate file (\texttt{summary\_files/inconsistent\_comment\_metadata}) with a list of annotated \texttt{Comment\_Id} with respective inconsistencies, followed by a column indicating the issue. 

Approximately $1,000$ article identifiers referenced in the comments files do not have corresponding entries in the articles' metadata. These cases are predominantly from older discussions, and the majority of them are inaccessible on the DerStandard website. We identify and list all $382,167$ comments ($0.5\%$ of the total amount of comments) belonging to these problematic article identifiers within the category of \textit{Article deleted}.

In some cases, the number of votes recorded in the votes files for a given comment is one higher than the number of votes reported in the comments files. This discrepancy affects only a small subset of comments and is not systematically associated with deleted authors, thread structure, or specific articles. It appears to result from minor inconsistencies in how vote totals were recorded on the platform. To resolve this, we replaced the vote counts in the comments files with the corresponding aggregated values from the votes files.

Finally, we observed a small number of comments ($1,208$) with timestamps that precede the timestamp of the article they are linked to. This issue is also visible on the DerStandard website and may stem from article republication, delayed timestamping, or legacy content migration. One such example is available at \url{https://www.derstandard.at/story/3222966/kosovo-unabhaengigkeit-und-separatistische-bewegungen}. These rare cases may slightly affect the computation of lifetime estimates of a very small subset of users (i.e. less than $0.15\%$ of the $247,863$ users) or temporal dynamics and should be considered carefully. We list all comments with inconsistent time stamps within the category of \textit{Inconsistent timestamp}.

\section*{Usage Notes}

As noted in the main text, we provide a separate file listing all comment identifiers associated with potential data inconsistencies. These include: (1) comments referencing article identifiers that are no longer available in the articles' metadata (nor accessible on the DerStandard website), and (2) comments with timestamps preceding those of their associated articles. While these cases represent a small fraction of the data, they may affect specific analyses, such as estimating user lifespans. We recommend that users assess whether these entries impact their workflow and, if necessary, exclude the small number of affected rows from their analysis.

Detailed instructions for downloading, decompressing, and using the dataset are included in the main README file within the repository. We strongly recommend that users consult this file before accessing the data. For example, the comment embedding vectors are split into multiple parts to avoid exceeding file size limits ($2.5$ GB per file). These files must be concatenated before decompression.

Finally, researchers using this dataset should ensure that their analyses comply with applicable ethical standards for computational social science research, including avoiding attempts to re-identify individuals or infer sensitive personal attributes from the anonymized data.

\section*{Data Availability}

The dataset described in this work is publicly available in the BSC Dataverse repository under the title “A Decade of News Forum Interactions: Threaded Conversations, Signed Votes, and Topical Tags”~\cite{RVQGUG_2026}. The dataset can be accessed via its DOI: \url{https://doi.org/10.82201/RVQGUG}. It is distributed under a CC-BY license and includes all data files, summary metadata, and example scripts required to access and process the dataset.

\section*{Code Availability}

Supporting code and example scripts to process the data can be found in the dataset folder \texttt{Scripts}~\cite{RVQGUG_2026}, including two scripts demonstrating how to scrape forum discussions and votes from the platform at the time data was obtained. Example code to join and decompress the embedding files from the repository can be found in the README file.


\section*{Author Contributions}

EF: Conceptualization, Methodology, Data Curation, Visualization, Technical Validations, Writing - Original Draft; Writing - Review and Editing. MP: Conceptualization, Methodology - Web Scraping, Methodology - Text Embeddings, Writing - Original Draft, Writing - Review and Editing. AK, VG: Conceptualization, Writing - Review and Editing.  All of the authors approved the final manuscript for submission.

\section*{Competing Interests}
Not applicable.

\section*{Data Protection and Ethical Approval}

The dataset was screened and approved by the Data Protection Office at the Barcelona Supercomputing Center. Following an internal review, it was confirmed that the General Data Protection Regulation (GDPR) does not apply to this dataset, as the data have been properly anonymized.

Although the dataset is derived from publicly accessible forum discussions on the DerStandard platform, ethical considerations remain important when sharing and analyzing large-scale online interaction data. Users contributing to online news discussions may not anticipate that their comments and interactions could later be aggregated and analyzed at scale for research purposes. To mitigate potential risks to user privacy, we do not release the raw text of comments and instead provide pre-computed embedding representations together with structural interaction data. Persistent identifiers have been anonymized through salted cryptographic hashing to prevent linkage to platform usernames while preserving relational structure within the dataset.

Despite these safeguards, a potential re-identification risk may exist if metadata were manually cross-referenced against the live website. We have taken reasonable technical measures to minimize this possibility. For example, article URLs have been removed from the public release to further reduce the possibility of indirect linkage to live content. 

Researchers using this dataset should therefore refrain from any attempts to re-identify individuals or to link anonymized records to real-world identities. Analyses should focus on aggregate patterns of interaction, communication dynamics, or structural properties of discussions rather than individual-level profiling.

\section*{Acknowledgements}

We thank Elena Candellone, Darja Cvetković, Özgür Togay, David March, Timur Naushirvanov, Simon D. Lindner, and Elena G. de Lamo for testing the dataset during the Winter Workshop on Complex Systems 2025 and providing valuable feedback on its integrity and usability. We also thank the BSC Dataverse Data Stewards, Adrian Carrascosa Lopez and Felipe Leonardo Gomez Cortes, for their support and guidance in publishing the dataset in accordance with FAIR principles.

\section*{Funding}

E.F. and V.G. acknowledge support from the project CNS2022-136178 financed by MCIN\slash AEI\slash 10.13039\slash 501100011033 and by the European Union Next Generation EU\slash PRTR.

This work is part of Maria de Maeztu Units of Excellence Programme CEX2021-001195-M, funded by MICIU/AEI /10.13039/501100011033.


\begin{thebibliography}{10}

\bibitem{murtfeldt2024rip}
R.~Murtfeldt, S.~Paik, N.~Alterman, I.~Kahveci, and J.~D. West, ``Rip twitter api: A eulogy to its vast research contributions,'' {\em arXiv preprint arXiv:2404.07340}, 2024.
\newblock arXiv:2404.07340.

\bibitem{bisbee2025vibes}
J.~Bisbee and K.~Munger, ``The vibes are off: did {Elon Musk} push academics off twitter?,'' {\em PS: Political Science \& Politics}, vol.~58, no.~1, pp.~139--146, 2025.
\newblock \url{https://doi.org/10.1017/S1049096524000416}.

\bibitem{naab2020comments}
T.~K. Naab, D.~Heinbach, M.~Ziegele, and M.-T. Grasberger, ``Comments and credibility: how critical user comments decrease perceived news article credibility,'' {\em Journalism studies}, vol.~21, no.~6, pp.~783--801, 2020.
\newblock \url{https://doi.org/10.1080/1461670X.2020.1724181}.

\bibitem{prochazka2018effects}
F.~Prochazka, P.~Weber, and W.~Schweiger, ``Effects of civility and reasoning in user comments on perceived journalistic quality,'' {\em Journalism studies}, vol.~19, no.~1, pp.~62--78, 2018.
\newblock \url{https://doi.org/10.1080/1461670X.2016.1161497}.

\bibitem{ziegele2014creates}
M.~Ziegele, T.~Breiner, and O.~Quiring, ``What creates interactivity in online news discussions? an exploratory analysis of discussion factors in user comments on news items,'' {\em Journal of Communication}, vol.~64, no.~6, pp.~1111--1138, 2014.
\newblock \url{https://doi.org/10.1111/jcom.12123}.

\bibitem{toepfl2015public}
F.~Toepfl and E.~Piwoni, ``Public spheres in interaction: Comment sections of news websites as counterpublic spaces,'' {\em Journal of Communication}, vol.~65, no.~3, pp.~465--488, 2015.
\newblock \url{https://doi.org/10.1111/jcom.12156}.

\bibitem{bacaksizlar2023group}
N.~G. Bacaksizlar~Turbic and M.~Galesic, ``Group threat, political extremity, and collective dynamics in online discussions,'' {\em Scientific Reports}, vol.~13, no.~1, p.~2206, 2023.
\newblock \url{https://doi.org/10.1038/s41598-023-28569-1}.

\bibitem{ha2025dynamics}
S.~Ha, H.~Olsson, K.~Jaksic, M.~Pellert, and M.~Galesic, ``Dynamics of collective minds in online communities,'' {\em arXiv preprint arXiv:2504.08152}, 2025.
\newblock arXiv:2504.08152.

\bibitem{russmann2020news}
U.~Russmann and A.~Hess, ``News consumption and trust in online and social media: An in-depth qualitative study of young adults in austria,'' {\em International Journal of Communication}, vol.~14, pp.~18--18, 2020.

\bibitem{russmann2025management}
U.~Russmann and A.~Hess, ``The management of uncivil and hateful user comments in austrian news media,'' {\em Journalism Practice}, vol.~19, no.~2, pp.~427--446, 2025.
\newblock \url{https://doi.org/10.1080/17512786.2023.2189152}.

\bibitem{schabus-skowron-2018-academic}
D.~Schabus and M.~Skowron, ``Academic-industrial perspective on the development and deployment of a moderation system for a newspaper website,'' in {\em Proceedings of the Eleventh International Conference on Language Resources and Evaluation ({{LREC}} 2018)}, (Miyazaki, Japan), European Language Resources Association (ELRA).

\bibitem{niederkrotenthaler2022mental}
T.~Niederkrotenthaler, Z.~Laido, S.~Kirchner, M.~Braun, H.~Metzler, T.~Waldh{\"o}r, M.~J. Strauss, D.~Garcia, and B.~Till, ``Mental health over nine months during the sars-cov2 pandemic: Representative cross-sectional survey in twelve waves between april and december 2020 in austria,'' {\em Journal of affective disorders}, vol.~296, pp.~49--58, 2022.
\newblock \url{https://doi.org/10.1016/j.jad.2021.08.153}.

\bibitem{pellertValidatingDailySocial2022}
M.~Pellert, H.~Metzler, M.~Matzenberger, and D.~Garcia, ``Validating daily social media macroscopes of emotions,'' {\em Scientific Reports}, vol.~12, p.~11236, Dec. 2022.
\newblock \url{https://doi.org/10.1038/s41598-022-14579-y}.

\bibitem{candellone2025negative}
E.~Candellone, S.-A. Babul, {\"O}.~Togay, A.~Bovet, and J.~Garcia-Bernardo, ``Negative ties highlight hidden extremes in social media polarization,'' {\em Network Science}, vol.~13, p.~e10, 2025.
\newblock \url{https://doi.org/10.1017/nws.2025.10006}.

\bibitem{barbera2015tweeting}
P.~Barber{\'a}, J.~T. Jost, J.~Nagler, J.~A. Tucker, and R.~Bonneau, ``Tweeting from left to right: Is online political communication more than an echo chamber?,'' {\em Psychological science}, vol.~26, no.~10, pp.~1531--1542, 2015.
\newblock \url{https://doi.org/10.1177/0956797615594620}.

\bibitem{aragon2017thread}
P.~Arag{\'o}n, V.~G{\'o}mez, and A.~Kaltenbrunner, ``To thread or not to thread: The impact of conversation threading on online discussion,'' in {\em Proceedings of the International AAAI Conference on Web and social media}, vol.~11, pp.~12--21, 2017.
\newblock \url{https://doi.org/10.1609/icwsm.v11i1.14880}.

\bibitem{kaltenbrunner2007description}
A.~Kaltenbrunner, V.~G{\'o}mez, and V.~L{\'o}pez, ``Description and prediction of slashdot activity,'' in {\em 2007 Latin American Web Conference (LA-WEB 2007)}, pp.~57--66, IEEE, 2007.
\newblock \url{https://doi.org/10.1109/LA-Web.2007.21}.

\bibitem{maniuBuildingSignedNetwork2011}
S.~Maniu, B.~Cautis, and T.~Abdessalem, ``Building a signed network from interactions in {{Wikipedia}},'' in {\em Databases and {{Social Networks}} on - {{DBSocial}} '11}, (Athens, Greece), pp.~19--24, ACM Press, 2011.
\newblock \url{https://doi.org/10.1145/1996413.1996417}.

\bibitem{pougue2021debagreement}
J.~Pougu{\'e}-Biyong, V.~Semenova, A.~Matton, R.~Han, A.~Kim, R.~Lambiotte, and D.~Farmer, ``Debagreement: A comment-reply dataset for (dis) agreement detection in online debates,'' in {\em Thirty-fifth conference on neural information processing systems datasets and benchmarks track (round 2)}, 2021.

\bibitem{baumgartner2019comparative}
F.~R. Baumgartner, C.~Breunig, and E.~Grossman, ``The comparative agendas project: Intellectual roots and current developments,'' 2019.

\bibitem{parlamint}
T.~Erjavec, M.~Kopp, T.~Kuzman~Punger{\v s}ek, N.~Ljube{\v s}i{\'c}, M.~Ogrodniczuk, P.~Osenova, M.~Agirrezabal, T.~Agnoloni, J.~Aires, M.~Albini, J.~Alkorta, I.~Antiba-Cartazo, E.~Arrieta, M.~Barcala, D.~Bardanca, S.~Barkarson, R.~Bartolini, R.~Battistoni, N.~Bel, M.~d.~M. Bonet~Ramos, M.~Calzada~P{\'e}rez, A.~Cardoso, {\c C}.~{\c C}{\"o}ltekin, M.~Coole, R.~Darģis, R.~de~Libano, G.~Depoorter, S.~Diwersy, R.~Dod{\'e}, K.~Fernandez, E.~Fern{\'a}ndez~Rei, F.~Frontini, M.~Garcia, N.~Garc{\'{\i}}a~D{\'{\i}}az, P.~Garc{\'{\i}}a~Louzao, M.~Gavriilidou, D.~Gkoumas, I.~Grigorov, V.~Grigorova, D.~Haltrup~Hansen, M.~Iruskieta, J.~Jarlbrink, K.~Jelencsik-M{\'a}tyus, B.~Jongejan, N.~Kahusk, M.~Kirnbauer, A.~Kryvenko, N.~Ligeti-Nagy, G.~Luxardo, C.~Magari{\~n}os, M.~Magnusson, C.~Marchetti, M.~Marx, K.~Meden, A.~Mendes, M.~Mochtak, M.~M{\"o}lder, S.~Montemagni, C.~Navarretta, B.~Nito{\'n}, F.~M. Nor{\'e}n, A.~Nwadukwe, M.~Ojster{\v s}ek, A.~Pan{\v c}ur, V.~Papavassiliou, R.~Pereira, M.~P{\'e}rez~Lago, S.~Piperidis,
  H.~Pirker, M.~Pisani, H.~v.~d. Pol, P.~Prokopidis, V.~Quochi, P.~Rayson, X.~L. Regueira, A.~Rii, M.~Rudolf, M.~Ruisi, P.~Rupnik, D.~Schopper, K.~Simov, L.~Sinikallio, J.~Skubic, L.~M. Tungland, J.~Tuominen, R.~van Heusden, Z.~Varga, M.~V{\'a}zquez~Abu{\'{\i}}n, G.~Venturi, A.~Vidal~Migu{\'e}ns, K.~Vider, A.~Vivel~Couso, A.~I. Vladu, T.~Wissik, V.~Yrj{\"a}n{\"a}inen, R.~Zevallos, and D.~Fi{\v s}er, ``Multilingual comparable corpora of parliamentary debates {ParlaMint} 5.0,'' 2025.
\newblock Slovenian language resource repository {CLARIN}.{SI}. \url{http://hdl.handle.net/11356/2004}.

\bibitem{rovny2025ches}
J.~Rovny, R.~Bakker, L.~Hooghe, S.~Jolly, G.~Marks, J.~Polk, J.~Rovny, M.~Steenbergen, and M.~A. Vachudova, ``The 2024 chapel hill expert survey on political party positioning in europe: Twenty-five years of party positional data,'' {\em Electoral Studies}, vol.~97, October 2025.
\newblock \url{https://doi.org/10.1016/j.electstud.2025.102981}.

\bibitem{P5YJ0O_2020}
B.~Kittel, S.~Kritzinger, H.~Boomgaarden, B.~Prainsack, J.-M. Eberl, F.~Kalleitner, N.~S. Lebernegg, J.~Partheymüller, C.~Plescia, D.~W. Schiestl, and L.~Schlogl, ``{Austrian Corona Panel Project (OA edition)},'' 2020.
\newblock AUSSDA. \url{https://doi.org/10.11587/P5YJ0O}.

\bibitem{schabusOneMillionPosts2017}
D.~Schabus, M.~Skowron, and M.~Trapp, ``One {{Million Posts}}: {{A Data Set}} of {{German Online Discussions}},'' in {\em Proceedings of the 40th {{International ACM SIGIR Conference}} on {{Research}} and {{Development}} in {{Information Retrieval}}}, pp.~1241--1244, ACM, 2017.

\bibitem{krenn2024germs}
B.~Krenn, J.~Petrak, M.~Kubina, and C.~Burger, ``Germs-at: A sexism/misogyny dataset of forum comments from an austrian online newspaper,'' in {\em Proceedings of the 2024 Joint International Conference on Computational Linguistics, Language Resources and Evaluation (LREC-COLING 2024)}, pp.~7728--7739, 2024.

\bibitem{pellertDashboardSentimentAustrian2020a}
M.~Pellert, J.~Lasser, H.~Metzler, and D.~Garcia, ``Dashboard of {{Sentiment}} in {{Austrian Social Media During COVID-19}},'' {\em Frontiers in Big Data}, vol.~3, 2020.
\newblock \url{https://doi.org/10.3389/fdata.2020.00032}.

\bibitem{10.1093/pnasnexus/pgae276}
E.~Fraxanet, M.~Pellert, S.~Schweighofer, V.~Gómez, and D.~Garcia, ``Unpacking polarization: Antagonism and alignment in signed networks of online interaction,'' {\em PNAS Nexus}, vol.~3, p.~pgae276, 07 2024.
\newblock \url{https://doi.org/10.1093/pnasnexus/pgae276}.

\bibitem{aumasson2013blake2}
J.-P. Aumasson, S.~Neves, Z.~Wilcox-O’Hearn, and C.~Winnerlein, ``Blake2: simpler, smaller, fast as md5,'' in {\em International Conference on Applied Cryptography and Network Security}, pp.~119--135, Springer, 2013.
\newblock \url{https://doi.org/10.1007/978-3-642-38980-1_8}.

\bibitem{morris2023text}
J.~Morris, V.~Kuleshov, V.~Shmatikov, and A.~M. Rush, ``Text embeddings reveal (almost) as much as text,'' in {\em Proceedings of the 2023 Conference on Empirical Methods in Natural Language Processing}, pp.~12448--12460, 2023.
\newblock \url{https://doi.org/10.18653/v1/2023.emnlp-main.765}.

\bibitem{jha2025harnessing}
R.~Jha, C.~Zhang, V.~Shmatikov, and J.~X. Morris, ``Harnessing the universal geometry of embeddings,'' {\em arXiv preprint arXiv:2505.12540}, 2025.
\newblock arXiv:2505.12540.

\bibitem{huang2024transferable}
Y.-H. Huang, Y.~Tsai, H.~Hsiao, H.-Y. Lin, and S.-D. Lin, ``Transferable embedding inversion attack: Uncovering privacy risks in text embeddings without model queries,'' in {\em Proceedings of the 62nd Annual Meeting of the Association for Computational Linguistics (Volume 1: Long Papers)}, pp.~4193--4205, 2024.
\newblock \url{https://doi.org/10.18653/v1/2024.acl-long.230}.

\bibitem{hayet2022invernet}
I.~Hayet, Z.~Yao, and B.~Luo, ``Invernet: An inversion attack framework to infer fine-tuning datasets through word embeddings,'' in {\em Findings of the Association for Computational Linguistics: EMNLP 2022}, pp.~5009--5018, 2022.
\newblock \url{https://doi.org/10.18653/v1/2022.findings-emnlp.368}.

\bibitem{li-etal-2023-sentence}
H.~Li, M.~Xu, and Y.~Song, ``Sentence embedding leaks more information than you expect: Generative embedding inversion attack to recover the whole sentence,'' in {\em Findings of the Association for Computational Linguistics: ACL 2023}, pp.~14022--14040, Association for Computational Linguistics, 2023.
\newblock \url{https://doi.org/10.18653/v1/2023.findings-acl.881}.

\bibitem{hu2025kalm}
X.~Hu, Z.~Shan, X.~Zhao, Z.~Sun, Z.~Liu, D.~Li, S.~Ye, X.~Wei, Q.~Chen, B.~Hu, {\em et~al.}, ``Kalm-embedding: Superior training data brings a stronger embedding model,'' {\em arXiv preprint arXiv:2501.01028}, 2025.
\newblock arXiv:2501.01028.

\bibitem{muennighoffMTEBMassiveText2023}
N.~Muennighoff, N.~Tazi, L.~Magne, and N.~Reimers, ``{{MTEB}}: {{Massive Text Embedding Benchmark}},'' in {\em Proceedings of the 17th {{Conference}} of the {{European Chapter}} of the {{Association}} for {{Computational Linguistics}}}, (Dubrovnik, Croatia), pp.~2014--2037, Association for Computational Linguistics, May 2023.
\newblock \url{https://doi.org/10.18653/v1/2023.eacl-main.148}.

\bibitem{RVQGUG_2026}
E.~Fraxanet, V.~Gómez, A.~Kaltenbrunner, and M.~Pellert, ``{(Dataset) A Decade of News Forum Interactions: Threaded Conversations, Signed Votes, and Topical Tags},'' 2026.
\newblock BSC Dataverse. \url{https://doi.org/10.82201/RVQGUG}.

\end{thebibliography}
\end{document}